\documentclass[epj]{webofc}
\usepackage[utf8]{inputenc}
\usepackage[varg]{txfonts}   
\usepackage{booktabs}
\usepackage{xcolor}
\def\LL{\left\langle}	
\def\RR{\right\rangle}	
\def\LP{\left(}		
\def\RP{\right)}	
\def\LS{\left[}	
\def\RS{\right]}	

\newcommand{\BE}{\begin{displaymath}}
\newcommand{\EE}{\end{displaymath}}
\newcommand{\EL}{\nonumber\\}

\newcommand{\ie}{{\em i.e.\ }}

\def\BEQ{\begin{equation}}
\def\ENQ{\end{equation}}
\def\BEA{\begin{eqnarray}}
\def\EEA{\end{eqnarray}}
\def\ENA{\nonumber\end{eqnarray}}
\def\BEI{\begin{itemize}}
\def\ENI{\end{itemize}}

\def\MBs{M_{B_s}}
\def\fpar{f_\parallel}
\def\fperp{f_\perp}

\usepackage{bm}
\definecolor{darkred}{rgb}{0.4,0.0,0.0}
\definecolor{darkgreen}{rgb}{0.0,0.4,0.0}
\definecolor{darkblue}{rgb}{0.0,0.0,0.4}
\usepackage[bookmarks,linktocpage,colorlinks,
    linkcolor = darkred,
    urlcolor  = darkblue,
    citecolor = darkgreen]{hyperref}
\usepackage{subfigure}
\wocname{EPJ Web of Conferences}
\woctitle{Lattice2017}
\begin{document}
%
\selectlanguage{english}
\title{%
\boldmath $B_s \to K \ell\nu$ form factors with 2+1 flavors
}
\author{%
\firstname{Yuzhi} \lastname{Liu}\inst{1} \and
\firstname{Jon A.} \lastname{Bailey}\inst{2} \and
\firstname{A.} \lastname{Bazavov}\inst{3} \and
\firstname{C.} \lastname{Bernard}\inst{4} \and
\firstname{C. M.} \lastname{Bouchard}\inst{5} \and
\firstname{C.} \lastname{DeTar}\inst{6} \and
\firstname{Daping} \lastname{Du}\inst{7} \and
\firstname{A. X.} \lastname{El-Khadra}\inst{8,9} \and
\firstname{E. D.} \lastname{Freeland}\inst{10} \and
\firstname{E.} \lastname{G\'amiz}\inst{11} \and
\firstname{Z.} \lastname{Gelzer}\inst{8,12} \and
\firstname{Steven} \lastname{Gottlieb}\inst{1}\fnsep\thanks{Speaker, \email{sg@indiana.edu} 
}
 \and
\firstname{U. M.} \lastname{Heller}\inst{13} \and
\firstname{A. S.} \lastname{Kronfeld}\inst{9,14} \and
\firstname{J.} \lastname{Laiho}\inst{7} \and
\firstname{P. B.} \lastname{Mackenzie}\inst{9} \and
\firstname{Y.} \lastname{Meurice}\inst{12} \and
\firstname{E. T.} \lastname{Neil}\inst{15,16} \and
\firstname{J. N.} \lastname{Simone}\inst{9} \and
\firstname{R.} \lastname{Sugar}\inst{17} \and
\firstname{D.} \lastname{Toussaint}\inst{18} \and
\firstname{R. S.} \lastname{Van de Water}\inst{9} \and
\firstname{Ran} \lastname{Zhou}\inst{9}
}
\institute{%
Department of Physics, Indiana University, Bloomington, Indiana 47405, USA
\and
Department of Physics and Astronomy, Seoul National University, Seoul, South Korea
\and
Department of Computational Mathematics, Science and Engineering and Department of Physics
and Astronomy, Michigan State University, East Lansing, Michigan, 48824, USA
\and
Department of Physics, Washington University, St.~Louis, Missouri, 63130, USA
\and
School  of  Physics  and  Astronomy,  University  of  Glasgow,  Glasgow,  G12  8QQ,  UK
\and
Department of Physics and Astronomy, University of Utah, Salt Lake City, Utah 84112, USA
\and
Department of Physics, Syracuse University, Syracuse, New York 13244, USA
\and
Department of Physics, University of Illinois, Urbana, Illinois 61801, USA
\and
Fermi National Accelerator Laboratory, Batavia, Illinois 60510, USA
\and
School of the Art Institute of Chicago, Chicago, Illinois 60603, USA
\and
CAFPE and Departamento de F\'{\i}sica Te\'orica y del Cosmos, Universidad de Granada, Granada, Spain
\and
Department of Physics and Astronomy, University of Iowa, Iowa City, IA 52242, USA
\and
American Physical Society, Ridge, New York 11961, USA
\and
Institute for Advanced Study, Technische Universit\"at M\"unchen, D-85748 Garching, Germany
\and
Department of Physics, University of Colorado, Boulder, CO 80309, USA
\and
RIKEN-BNL Research Center, Brookhaven National Laboratory, Upton, NY 11973, USA
\and
Department of Physics, University of California, Santa Barbara, California 93106, USA
\and
Department of Physics, University of Arizona, Tucson, Arizona 85721, USA
}
\abstract{%
Using the MILC 2+1 flavor asqtad quark action ensembles, we are calculating
the form factors $f_0$ and $f_+$ for the semileptonic
$B_s \rightarrow K \ell\nu$ decay.  A total of six ensembles with lattice
spacing from $\approx0.12$ to 0.06 fm are being used.  At the coarsest
and finest lattice spacings, the light quark mass $m'_l$ is one-tenth the
strange quark mass $m'_s$.  
At the intermediate lattice spacing, the ratio $m'_l/m'_s$
ranges from 0.05 to 0.2.  The valence $b$ quark is treated using the
Sheikholeslami-Wohlert Wilson-clover action with the Fermilab interpretation.
The other valence quarks use the asqtad action.  When combined with (future)
measurements from the LHCb and Belle II experiments, these calculations will provide 
an alternate determination of the CKM matrix element $|V_{ub}|$.

}
\maketitle
\vfill
\section{Introduction}\label{sec-1}

\begin{figure}[bp]
   \centering
             {\includegraphics[width=0.475\textwidth,clip]{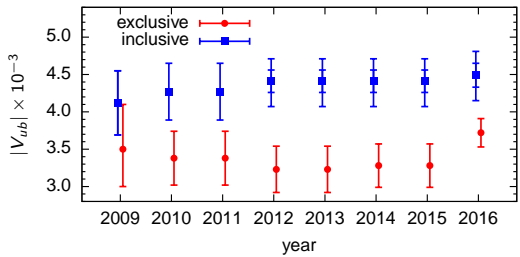}}
   \caption{History of the tension between determination of $|V_{ub}|$
from exclusive and inclusive decays~\cite{Olive:2016xmw}.}
   \label{fig:exclvsincl}
\end{figure}

\begin{figure}[tp]
   \centering
             \includegraphics[width=0.475\textwidth,clip]{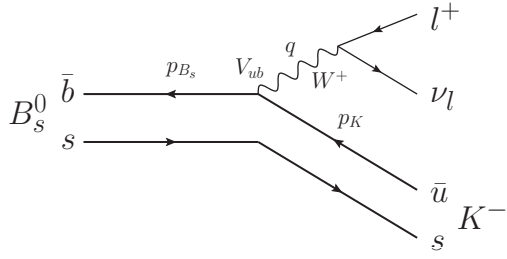}
   \caption{Feynman diagram for the decay $B_s \to K^- \ell^+ \nu$ without
any of the QCD corrections.}
   \label{fig:feynmandiagram}
\end{figure}

The Cabibbo-Kobayashi-Masakawa (CKM) matrix describes weak interaction mixing
of quarks in the Standard Model of elementary particle and nuclear physics.
The elements of the matrix are fundamental parameters
of the Standard Model.  If the CKM matrix is not unitary, or if independent
determinations of a particular matrix element from different decays do
not agree, that provides evidence of new physics beyond the Standard Model.
The element $|V_{ub}|$ is an important avenue to search for new physics.
There is a long standing tension between its value determined from inclusive 
and exclusive decays.  (See Figure~\ref{fig:exclvsincl}.)
The exclusive decay $B \to \pi^- \ell^+
\nu$~\cite{Lattice:2015tia} has been used to determine $|V_{ub}|$.  
The theory error from lattice calculations are smaller for the process at hand,
$B_s \to K^- \ell^+ \nu$, because the spectator
quark is strange, rather than an up or down quark.  Experimental measurements
of this decay will be available from LHCb and Belle II.  
Figure~\ref{fig:feynmandiagram} shows a Feynman diagram of the decay
without any of the QCD corrections that connect the valence quarks.

In this work, we use the 2+1 flavor MILC asqtad ensembles 
\cite{Bernard:2001av,Aubin:2004wf,Bazavov:2009bb}, asqtad valence
light and strange quarks, and clover quarks with the Fermilab interpretation
for the $b$ quark~\cite{PhysRevD.55.3933}.  The decay has also been studied by
HPQCD~\cite{Bouchard:2014ypa} using MILC asqtad ensembles with 
HISQ light valence quarks and
an NRQCD $b$ quark.  The RBC and UKQCD Collaborations~\cite{Flynn:2015mha} have
used 2+1 flavor domain-wall dynamical quark ensembles, domain-wall valence
light quarks and a relativistic heavy quark action for the $b$
quark.

\section{Matrix elements and form factors}\label{sec-2}
Lattice QCD allows us to compute the hadronic matrix elements that are needed
to calculate the decay amplitudes.  The matrix elements can be expressed
in terms of form factors in two ways:
\begin{eqnarray}
\LL K(p_K)|\bar{u}\gamma^\mu b|B_s(p_{B_s})\RR
&=&
\LP p^\mu_K + p^\mu_{B_s} -q^\mu\frac{\MBs^2 - M_K^2}{q^2} \RP f_+(q^2) +
q^\mu\frac{\MBs^2 - M_K^2}{q^2}f_0(q^2) \EL
&=&
\sqrt{2\MBs} \LS v^\mu \fpar(E_K) + p^\mu_{\perp}\fperp(E_K) \RS.
\label{eq:matrix_element_cont}
\end{eqnarray}
The initial $B_s$-meson 4-momentum is $p_{B_s}$, the final kaon 4-momentum
is $p_K$, and the 4-momentum transfer to the leptons is $q$.  Two form factors
appear on the RHS, either $f_+$ and $f_0$ or $\fpar$ and $\fperp$.
In the second expression, $v^\mu \equiv p_{B_s}^\mu/M_{B_s}$ is the 4-velocity 
of the $B_s$ meson and
$p_\perp^\mu\equiv p_K^\mu-(p_K\cdot v)v^\mu$ is the part of the kaon 
4-momentum orthogonal to $v$.
The vector form factor $f_+(q^2)$ and the scalar form factor $f_0(q^2)$
satisfy the kinematic constraint $f_+(0) = f_0(0)$.
The two sets of form factors are related by:
\begin{subequations}
\begin{eqnarray}
f_+(q^2) &=& \frac{1}{\sqrt{2M_{B_s}}}[\fpar(E_K) + (M_{B_s} - E_K) \fperp(E_K)],\\
f_0(q^2) &=& \frac{\sqrt{2M_{B_s}}}{M_{B_s}^2 - M_K^2} [(\MBs - E_K)\fpar(E_K)
+ (E_K^2 - M_K^2)\fperp(E_K)].
\end{eqnarray}
\label{eq:fp_f0_fpar_fperp}
\end{subequations}

In lattice QCD, it is convenient to calculate the 
second set of form factors on each ensemble by calculating appropriate 2-
and 3-point functions.  The former determine the $B_s$ meson mass, and 
the kaon meson mass and energy (as a function of the $K$ 3-momentum $\bm{p}_K$).
The 3-point functions determine the lattice form factors 
$\fpar^{\mathrm{lat}}$ and $\fperp^{\mathrm{lat}}$ at the corresponding
energies.  After matching the currents, 
we obtain the continuum $\fpar$ and $\fperp$ by performing a chiral-continuum
fit and extrapolating the fit to physical quark masses and the 
continuum (zero lattice spacing) limit.
The continuum form factors $f_+$ and $f_0$ are constructed from $\fpar$ and
$\fperp$ via Eqs.~(\ref{eq:fp_f0_fpar_fperp}) and extrapolated to the whole
kinematically allowed momentum transfer region using the
$z$-expansion~\cite{Bourrely:2008za,Lattice:2015tia}.

\section{Correlation functions}\label{sec-3}
To carry out this calculation we need a variety of 2- and 3-point
correlation functions.  We define them as:
\begin{subequations}
\begin{eqnarray}
C_2^{B_s}(t;0)  &=& \sum_{\bm{x}}
\langle
\mathcal{O}_{B_s}(t,\bm{x})
\mathcal{O}_{B_s}^\dagger(0,\bm{0})
\rangle ,  \\
C_2^{K}(t;\bm{p}_K)  &=& \sum_{\bm{x}}
\langle
\mathcal{O}_{K}(t,\bm{x})
\mathcal{O}_{K}^\dagger(0,\bm{0})
\rangle
e^{-i \bm{p}_K\cdot \bm{x}} , \\
C_{3,\mu}^{B_s \rightarrow K}(t,T;\bm{p}_K) &=& \sum_{\bm{x},\bm{y}}
\langle \mathcal{O}_{K} (0,\bm{0}) V^\mu(t,\bm{y})
\mathcal{O}_{B_s}^\dagger(T,\bm{x}) \rangle
e^{i \bm{p}_K\cdot \bm{y}},
\label{subeq:cont_3pt}
\end{eqnarray}
\label{eq:cont_2pt_3pt}
\end{subequations}
where $p_K$ is the kaon momentum and 
$ V^\mu$ is the lattice vector current. The continuum vector current
$\mathcal{V}^\mu \equiv \bar{u}\gamma^\mu b  = Z_{V_\mu} V^\mu$ 
is related to the lattice one by
a renormalization factor $ Z_{V_\mu}$, which is blinded until our results
are finalized to avoid any bias during the analysis.

The 2-point correlators are used to extract the lattice meson masses and
to verify the dispersion relation for the kaon.  They also determine the
overlaps of the lattice operators $\mathcal{O}_{B_s}$ and $\mathcal{O}_K$
with the $B_s$ and $K$ states, respectively.

\section{Lattice details}\label{sec-4}
We use six of the MILC 2+1-flavor asqtad ensembles, with lattice spacings of
$\approx 0.12$, 0.09, and 0.06 fm.  For each lattice spacing, we have dynamical
sea quarks with mass ratio $m'_l/m'_s=0.1$.
For $a\approx 0.09$ fm, we have three
additional values of  $m'_l/m'_s=0.2$, 0.15 and 0.05 to provide results for the
chiral extrapolation.  We use asqtad valence quarks.  The valence $u$
and $d$ quarks are taken to be degenerate, and their mass $am_l$ is the
same as the light sea quark mass $am'_l$ on the corresponding ensemble.
However, the valence $s$-quark mass $am_s$ is better tuned to match the
physical value than the dynamical $s$-quark mass.  This subset of the MILC
ensembles was chosen based on our experience studying $B \to \pi$~\cite{Lattice:2015tia} and
$B\to K$~\cite{Bailey:2015dka} semileptonic decays.

\begin{table}[thb]
  \small
  \centering
  \caption{Table of ensembles used and key parameters.  From left to right:
approximate lattice spacing; grid size; sea light and strange quark masses
in lattice units; valence  strange quark masses
in lattice units; number of configurations analyzed; number of different time
sources used on each ensemble; product of pion mass and spatial size.}
  \label{tab-1}
\begin{tabular}{c c c c c c c}
\toprule
$\approx a$(fm)   &$N_s^3 \times N_t$ & $am'_l$/$am'_s$ &
$am_s$ & $N_\text{config}$   & $N_\text{source}$
&$a M_{\pi} N_s$\\
\midrule
0.12  &     $24^3\times64$  &0.0050/0.0500   &    0.0336    &     2099    &     4  &3.8\\
\midrule
0.09  &     $28^3\times96$  &0.0062/0.031   &    0.0247    &     1931    &     4   &4.1\\
0.09  &     $32^3\times96$  &0.00465/0.031   &    0.0247    &     1015    &   8   &4.1\\
0.09  &     $40^3\times96$  &0.0031/0.031   &    0.0247    &     1015    &     8   &4.2\\
0.09  &     $64^3\times96$  &0.00155/0.031   &    0.0247    &     791    &    4   &4.8\\
\midrule
0.06  &     $64^3\times144$  &0.0018/0.018   &    0.0177    &     827    &      4   &4.3\\
\bottomrule
\end{tabular}
\end{table}
The 2-point correlators are fit to these functional forms:
\begin{subequations}
\begin{eqnarray}
C_2^{B_s}(t;0) &=&
\sum_{n=0}^{2N-1}
(-1)^{n(t+1)}
\frac{|\langle 0|\mathcal{O}_{B_s}|{B_s}^{(n)} \rangle|^2}{2M_{B_s}^{(n)}}
\left(e^{-M_{B_s}^{(n)}t} + e^{-M_{B_s}^{(n)}(N_t - t)} \right),
\label{eq:lat_2pt_Bs} \\
C_2^K(t;\bm{p}_K) &=&
\sum_{n=0}^{2N-1}
(-1)^{n(t+1)}
\frac{|\langle 0|\mathcal{O}_K|K^{(n)} \rangle|^2}{2E_K^{(n)}}
\left(e^{-E_K^{(n)}t} + e^{-E_K^{(n)}(N_t - t)} \right).
\label{eq:lat_2pt_Kaon}
\end{eqnarray}
\end{subequations}
We use $N=3$ in our fits, with prior central values for $n=0$ 
based on
effective masses.  We have set the prior widths widely enough to avoid
bias.  We fit over the $t$ range [$t_{\text{min}}$,$t_{\text{max}}$], with $t_{\text{min}}$
selected so that the fit has a good $p$-value and the ground state
energy is stable under variations in $t_{\text{min}}$.  We choose $t_{\text{max}}$ so
that the fractional error in the correlator is $<$3\%, thereby ignoring any noisy
tail at large $t$.  We use kaon 3-momentum up to $(1,1,1)\times 2\pi/N_s$
in lattice units, and have verified that the energy-momentum dispersion
relation is well satisfied.

The 3-point correlators are described by
\begin{eqnarray}
C_{3,\mu}^{B_s \rightarrow K}(t,T;\bm{p}_K) &=&
\sum_{m,n=0}^{2N-1}
(-1)^{m(t+1)}(-1)^{n(T-t-1)}
A^{\mu}_{mn} e^{-E_K^{(m)}t} e^{-M_{B_s}^{(n)}(T-t)},
\label{eq:lat_3pt}
\end{eqnarray}
where
\begin{eqnarray}
A_{mn}^{\mu} &=& \frac{\langle{0}|\mathcal{O}_K|{K^{(m)}}\rangle}{2E_K^{(m)}} 
\langle{K^{(m)}}|V^\mu|{{B_s}^{(n)}}\rangle\frac{\langle{{B_s}^{(n)}}|\mathcal{O}_{B_s}|{0}\rangle}{2M_{B_s}^{(n)}}.
\label{eq:Amn}
\end{eqnarray}                                                                

Since the energies and amplitudes are common to 2- and 3-point functions,
it is possible to fit them simultaneously.
An example of this fit for $a\approx 0.12$ fm can be found in 
Ref.~\cite{Liu:2013sya}, Figure 2.

\section{Chiral-continuum extrapolation}\label{sec-5}
Having extracted the lattice form factors on each ensemble for several
values of $E_K$, we are ready to perform the chiral-continuum fit.  We do this
using SU(2) heavy-meson rooted-staggered chiral perturbation
theory
(HMrS$\chi$P)~\cite{PhysRevD.73.014515,PhysRevD.76.014002,Bailey:2015dka}.  At
next to leading order, each form factor $f_P$ is fit to the form
\BEQ
f_{P, \mathrm{NLO}} =  f_{P}^{(0)} [
c^{0}_{P}(1+\delta f_{P, \text{logs}})+ c^{\text{1}}_{P}\chi_\text{l}+
c^{\text{2}}_{P}\chi_\text{h}+ c^{3}_{P}{\chi_E}+ c^{4}_{P}\chi_E^{2}+
c^5_{P}{\chi_a^{2}} ],
\label{eq:chiral_f_nlo}
\ENQ
where the leading order term $f^{(0)}_P$ is
\BEQ
\frac{1}{f_P}\frac{g_\pi}{E_K + \Delta_P^*}.
\ENQ
There is a pole determined by $\Delta_P^*$ which takes the form
\BEQ
\Delta_P^* = \frac{M_{B^*}^2 - M_{B_s}^2 - M_K^2}{2M_{B_s}}.
\label{eq:Delta_P}
\ENQ
We require $\fpar$ and $\fperp$ to have the same pole as $f_0$ and $f_+$,
respectively.  This is reasonable because, 
by Eq.~(\ref{eq:fp_f0_fpar_fperp}) $\fpar$ and $\fperp$ are dominated 
by contributions from  $f_0$ and $f_+$, respectively.
The vector meson (with $J^P = 1^-$) has been experimentally 
measured
~\cite{Olive:2016xmw}
as $M_{B^*} = 5324.65(25)~\mathrm{MeV}$.
The scalar $B^*$ meson (with $J^P = 0^+$) has not been observed experimentally,
but a lattice QCD calculation
~\cite{Gregory:2010gm}
suggests the mass difference between $0^+$ and $0^-$ states to be around 
$400~\mathrm{MeV}$, \ie $M_{B^*}(0^+) - M_B \approx 400~\mathrm{MeV}$.
The $J^P = 1^-$ pole is below the $B\pi$ production threshold, while the $0^+$
one is slightly above it, but still has a significant influence on the shape of the form factor.
The $c_P^i$ are coefficients of the corrections that depend on quark masses,
kaon energy, square of kaon energy, and square of lattice spacing.  They
are fit parameters.  Details of the chiral logarithms can be found in~\cite{Bailey:2015dka}.

\begin{figure}[tp]
   \centering
   \subfigure[$\fperp$ data and fit lines for each ensemble.  
    $\fpar$ and $\fperp$ are fit simultaneously.  
    The cyan band shows the continuum limit.]%
             {\includegraphics[width=0.475\textwidth,clip]{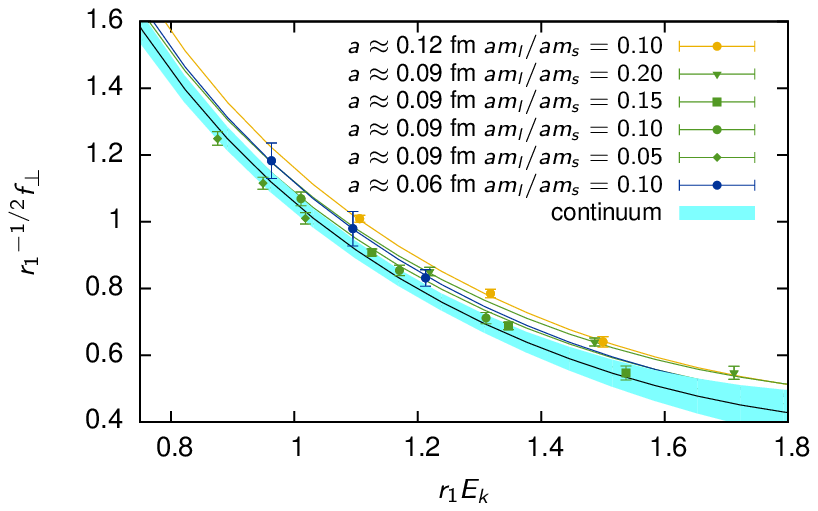}}\hfill
   \subfigure[$\fpar$ data and fit lines for each ensemble.]%
             {\includegraphics[width=0.475\textwidth,clip]{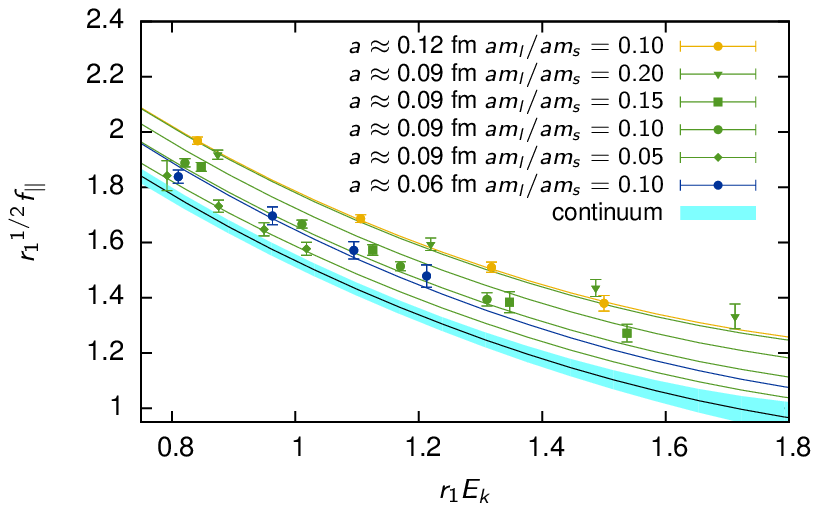}}
   \caption{Chiral-continuum fit to the lattice form factors. These form
factors are blinded, {\emph i.e.}, the person doing the analysis is given
the current renormalizations, but they have been multiplied by a blinding factor
only known to the person supplying the renormalizations.  Only after the
analysis is complete will the blinding factor be revealed so that 
the form factors can be properly normalized.}
   \label{fig:chiralcontinuum}
\end{figure}

For our central fit of $\fpar$ and $\fperp$ 
we allow additional NNLO analytic terms~\cite{Lattice:2015tia}, 
fitting both form factors simultaneously.
Figure \ref{fig:chiralcontinuum} shows the result of our fit.  
We note that $\chi^2/\textrm{dof} =0.89$ with
42 degrees of freedom corresponding to a $p$-value of 0.68.

\begin{figure}[tp]
   \centering
   \subfigure[Different sources of error for $f_+$ as a function of $q^2$.]%
             {\includegraphics[width=0.475\textwidth,clip]{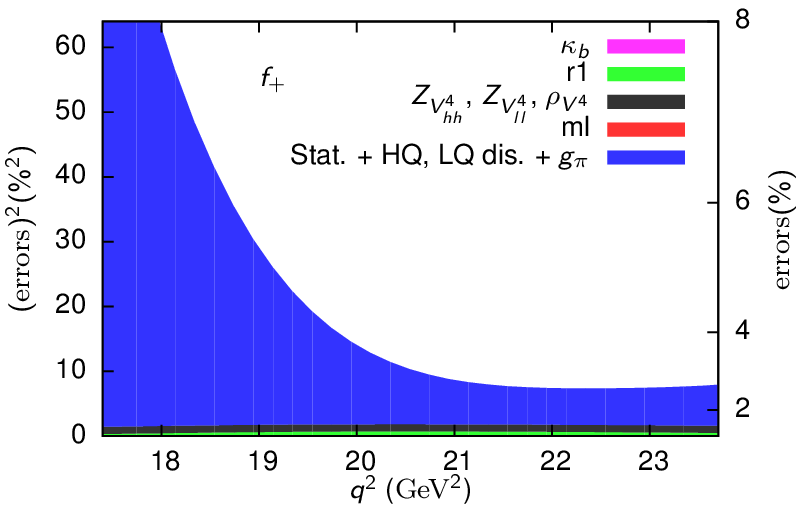}}\hfill
   \subfigure[Different sources of error for $f_0$ as a function of $q^2$.]%
             {\includegraphics[width=0.475\textwidth,clip]{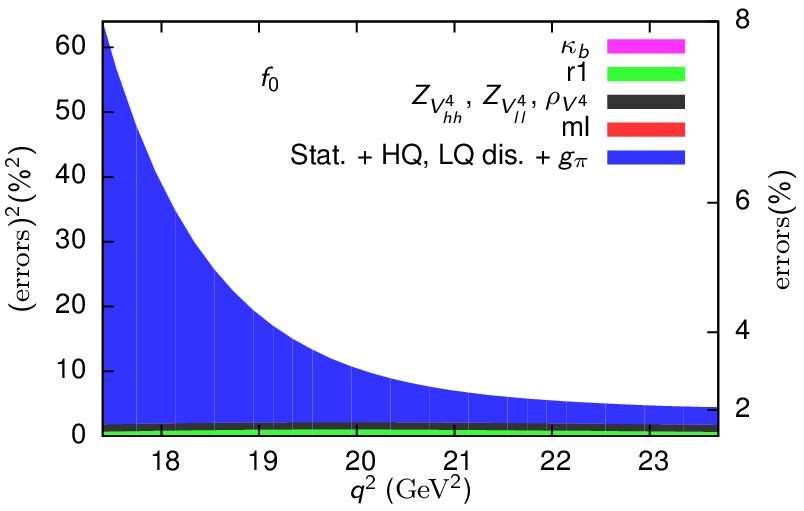}}
   \caption{Error budgets from statistical and systematic effects.}
   \label{fig:errorbudget}
\end{figure}

There are several other sources of systematic error that must also be taken
into account.  These include tuning of $\kappa_b$ needed to get the right
$b$-quark mass, possible mistuning of $am'_l$ from its physical value $am_l$,
uncertainty in the physical value of $r_1$~\cite{Bazavov:2011aa}, and the 
uncertainty in the renormalization of the vector current.
The quantity $r_1$ is related to the static potential and a variation
on the Sommer scale~\cite{Sommer:1993ce}.  It is dicussed extensively in
Ref.~\cite{Bazavov:2009bb}, Sec.~IV.B.
Discretization effects from the light and heavy quark actions and errors
in the coupling $g_\pi$ needed for the chiral logarithms are combined with
the statistical errors because they are parameters in the
chiral-continuum fit.  At this point, we use Eq.~(\ref{eq:fp_f0_fpar_fperp})
to convert from the lattice form factors to $f_+$ and $f_0$.  
In Fig.~\ref{fig:errorbudget}, we show our \emph{preliminary}
error budgets.  Note that the statistical errors dominate
the systematic errors, and they grow rapidly as $q^2$ decreases.

\begin{figure}[thb] 
  \centering
  \includegraphics[width=0.6\textwidth,clip]{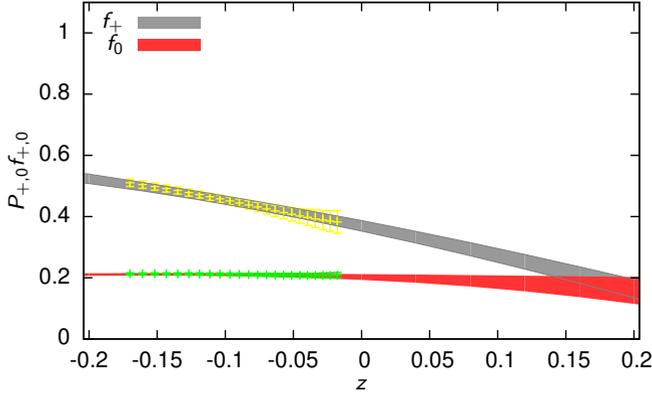}
  \caption{Blinded form factors as a function of $z$. The
region in which there is lattice data is shown with its errors in yellow
for $f_+$ and green for $f_0$.}
  \label{fig:zexpansion}
\end{figure}

\begin{figure}[thb] 
  \centering
  \includegraphics[width=0.6\textwidth,clip]{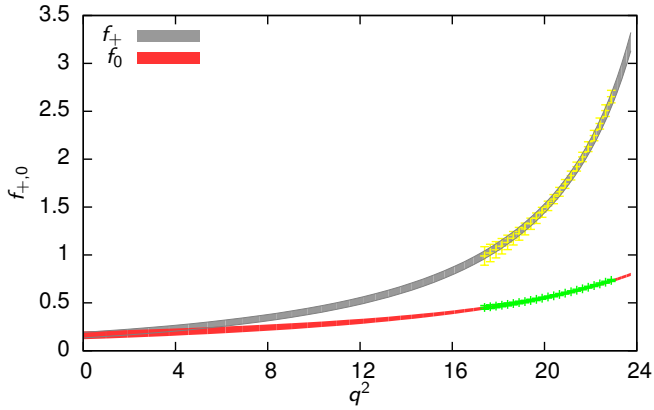}
  \caption{Blinded form factors shown in previous figure are now plotted \emph{vs.}
$q^2$.  Color scheme is the same as before.}
  \label{fig:preliminaryfinal}
\end{figure}

To reduce errors and extrapolate to $q^2 < 17 \textrm{GeV}^2$, we use the functional $z$-expansion
method described in Ref.~\cite{Lattice:2015tia}.
This avoids construction and fitting of synthetic
data.  For the $z$-expansion, we use the so-called BCL approach first
described in Ref.~\cite{Bourrely:2008za}.  
We fit $f_+$ and $f_0$ simultaneously 
keeping terms up to order $z^3$ without imposing the kinematic
constraint  $f_+(q^2 = 0) = f_0(q^2 = 0)$.  We see in Fig.~\ref{fig:zexpansion}
that this condition is well satisfied as $q^2=0$ corresponds to the maximum
value of $z$ in the figure.  We also note that the unitarity condition
$\sum_{m,n=0}^K B_{mn}b_m b_n \leq 1$ is well satisfied by our fit.
The sums are 0.160(30) for $f_+$ and 0.157(45) for $f_0$.
Imposition of constraints from heavy quark effective theory or kinematics
would only slightly reduce the error in the form factors at $q^2 = 0$.
Our  $z$-expansion fit has $\chi^2/dof = 0.82$ for 5 degrees of
freedom which corresponds to a $p$-value of 0.54.
We next reconstruct the form factors as functions of $q^2$.  Our 
preliminary result, for which the $Z$ factors for current renormalization
are still blinded, is shown in Fig.~\ref{fig:preliminaryfinal}.


\section{Summary}\label{sec-6}
This paper contains an update on our lattice QCD calculation of the form factors
$f_+$ and $f_0$ for the decay $B_s \to K \ell\nu$.  
Our results are still preliminary.  Once we finalize the systematic error
analysis, we will unblind the form factors, and compare them 
to previous results.
Before unblinding, we can only predict the shape of the decay distribution,
not its absolute magnitude.
Once we unblind, we
can use existing information about $|V_{ub}|$ to predict the $B_s$
differential decay rate.  Alternatively, once the decay distribution is
experimentally measured, our form factors can be used to infer
$|V_{ub}|$ from this decay.  
This may shed light on the current discrepancy between exclusive and inclusive
modes.

\bigskip
{\emph Acknowledgments:}
Computations for this work were carried out with resources provided by the USQCD Collaboration, the National
Energy Research Scientific Computing Center and the Argonne Leadership Computing Facility, which are funded
by the Office of Science of the U.S.\ Department of Energy; and with resources provided by the National
Institute for Computational Science and the Texas Advanced Computing Center, which are funded through the
National Science Foundation's Teragrid/XSEDE Program.  

This work was supported in part by the U.S.\ Department of Energy under grants
No.~DE-AC05-06OR23177 (B.C.),
No.~DE{-}SC0010120 (S.G.), 
No.~DE{-}SC0015655 (A.X.K.), 
No.~DE(-)SC0009998 (J.L.),
No.~DE(-)SC0010113 (Y.M.),
No.~DE{-}SC0010005 (E.T.N.), 
No.~DE-FG02-13ER41976 (D.T.),
by the U.S.\ National Science Foundation under grants
PHY14-17805~(D.D., J.L.), 
PHY12-12389~(Y.L.),
PHY14-14614 (C.D.), 
and PHY13-16748 and PHY16-20625 (R.S.); 
by the Fermilab Distinguished Scholars Program (A.X.K.);
by the German Excellence Initiative and the European Union Seventh Framework Program under grant agreement No.~291763 as well as the European Union's Marie Curie COFUND program (A.S.K.),
and by Spanish MINECO under grant No. FPA2013-47836-C3-1-P (E.G.).
Y.L. was partially supported by the Blue Waters PAID program.  The Blue 
Waters sustained-petascale computing project, which is supported by the 
National Science Foundation (awards OCI-0725070 and ACI-1238993) and the 
state of Illinois. Blue Waters is a joint effort of the 
University of Illinois at Urbana-Champaign and its National Center for 
Supercomputing Applications.

Fermilab is operated by Fermi Research Alliance, LLC, under Contract No.\ DE-AC02-07CH11359 with the United States Department of
Energy, Office of Science, Office of High Energy Physics.
The United States Government retains and the publisher, by accepting the article for publication, acknowledges that the United
States Government retains a non-exclusive, paid-up, irrevocable, world-wide license to publish or reproduce the published form of this manuscript, or allow others to do so, for United States Government purposes.


\bibliography{./lattice2017}

\begin{thebibliography}{16}

\bibitem{Lattice:2015tia}
J.A. Bailey et~al. (Fermilab Lattice, MILC), Phys. Rev. \textbf{D92}, 014024
  (2015), \texttt{1503.07839}

\bibitem{Bernard:2001av}
C.W. Bernard, T.~Burch, K.~Orginos, D.~Toussaint, T.A. DeGrand, C.E. Detar,
  S.~Datta, S.A. Gottlieb, U.M. Heller, R.~Sugar, Phys. Rev. \textbf{D64},
  054506 (2001), \texttt{hep-lat/0104002}

\bibitem{Aubin:2004wf}
C.~Aubin, C.~Bernard, C.~DeTar, J.~Osborn, S.~Gottlieb, E.B. Gregory,
  D.~Toussaint, U.M. Heller, J.E. Hetrick, R.~Sugar, Phys. Rev. \textbf{D70},
  094505 (2004), \texttt{hep-lat/0402030}

\bibitem{Bazavov:2009bb}
A.~Bazavov et~al. (MILC), Rev. Mod. Phys. \textbf{82}, 1349 (2010),
  \texttt{0903.3598}

\bibitem{PhysRevD.55.3933}
A.X. El-Khadra, A.S. Kronfeld, P.B. Mackenzie, Phys. Rev. D \textbf{55}, 3933
  (1997)

\bibitem{Bouchard:2014ypa}
C.M. Bouchard, G.P. Lepage, C.~Monahan, H.~Na, J.~Shigemitsu, Phys. Rev.
  \textbf{D90}, 054506 (2014), \texttt{1406.2279}

\bibitem{Flynn:2015mha}
J.M. Flynn, T.~Izubuchi, T.~Kawanai, C.~Lehner, A.~Soni, R.S. Van~de Water,
  O.~Witzel, Phys. Rev. \textbf{D91}, 074510 (2015), \texttt{1501.05373}

\bibitem{Olive:2016xmw}
C.~Patrignani et~al. (Particle Data Group), Chin. Phys. \textbf{C40}, 100001
  (2016)

\bibitem{Bourrely:2008za}
C.~Bourrely, I.~Caprini, L.~Lellouch, Phys. Rev. \textbf{D79}, 013008 (2009),
  [Erratum: Phys. Rev.D82,099902(2010)], \texttt{0807.2722}

\bibitem{Bailey:2015dka}
J.A. Bailey et~al., Phys. Rev. \textbf{D93}, 025026 (2016), \texttt{1509.06235}

\bibitem{Liu:2013sya}
Y.~Liu et~al., PoS \textbf{LATTICE2013}, 386 (2014), \texttt{1312.3197}

\bibitem{PhysRevD.73.014515}
C.~Aubin, C.~Bernard, Phys. Rev. D \textbf{73}, 014515 (2006)

\bibitem{PhysRevD.76.014002}
C.~Aubin, C.~Bernard, Phys. Rev. D \textbf{76}, 014002 (2007)

\bibitem{Gregory:2010gm}
E.B. Gregory et~al., Phys. Rev. \textbf{D83}, 014506 (2011), \texttt{1010.3848}

\bibitem{Bazavov:2011aa}
A.~Bazavov et~al. (Fermilab Lattice, MILC), Phys. Rev. \textbf{D85}, 114506
  (2012), \texttt{1112.3051}

\bibitem{Sommer:1993ce}
R.~Sommer, Nucl. Phys. \textbf{B411}, 839 (1994), \texttt{hep-lat/9310022}

\end{thebibliography}
\end{document}